\documentclass[12pt]{article}
\usepackage{amssymb}
\usepackage{amsmath}
\usepackage{graphicx,psfrag,epsf}
\usepackage{enumerate}
\usepackage{natbib}
\usepackage{tabularx}
\usepackage{url} 
\usepackage{tikz}
\usetikzlibrary{calc,arrows,positioning,decorations.pathreplacing}
\usepackage{bm}
\usepackage{float}
\usepackage{multirow}
\usepackage{threeparttable}
\usepackage{authblk}
\usepackage{booktabs}
\usepackage{dsfont}
\usepackage{tikz-network}
\usepackage{mathtools}
\newtheorem{prop}{Proposition}

\newtheorem{coro}{Corollary}

\newtheorem{theo}{Theorem}

\makeatletter
\renewcommand*\env@matrix[1][\arraystretch]{%
  \edef\arraystretch{#1}%
  \hskip -\arraycolsep
  \let\@ifnextchar\new@ifnextchar
  \array{*\c@MaxMatrixCols c}}
\makeatother

\begin{document}

\def\spacingset#1{\renewcommand{\baselinestretch}%
{#1}\small\normalsize} \spacingset{1}



  \title{\bf Neighbourhood Bootstrap for Respondent-Driven Sampling}
\author[1]{Mamadou Yauck\thanks{
  E-mail: \textit{mamadou.yauck@mcgill.ca}}}
\author[1]{Erica E. M. Moodie}
\author[2,3]{Herak Apelian}
\author[2,3]{Alain Fourmigue}
\author[5]{Daniel Grace}
\author[6]{Trevor A. Hart}
\author[2,4]{Gilles Lambert}
\author[1,2]{Joseph Cox}
\affil[1]{Department of Epidemiology, Biostatistics \& Occupational Health, McGill University, Montréal, QC, Canada}
\affil[2]{Direction régionale de santé publique de Montréal, CIUSSS Centre-Sud-de-l'Ile-de-Montréal, Montréal, QC, Canada }
\affil[3]{Research Institute of the McGill University Health Centre, Montréal, QC, Canada}
\affil[4]{Institut national de santé publique du Québec, Montréal, QC, Canada}
\affil[5]{Dalla Lana School of Public Health, University of Toronto}
\affil[6]{ Ryerson University, Toronto, Ontario, Canada}
\date{}
  \maketitle


\bigskip
\begin{abstract}
Respondent-Driven Sampling (RDS) is a form of link-tracing sampling, a sampling technique used for `hard-to-reach' populations that aims to leverage individuals' social relationships to reach potential participants. While the methodological focus has been restricted to the estimation of population proportions, there is a growing interest in the estimation of uncertainty for RDS as recent findings suggest that most variance estimators underestimate variability. Recently, \cite{baraff2016estimating} proposed the \textit{tree bootstrap} method based on resampling the RDS recruitment tree, and empirically showed that this method outperforms current bootstrap methods. However, some findings suggest that the tree bootstrap (severely) overestimates uncertainty. In this paper, we propose the \textit{neighbourhood} bootstrap method for quantifiying uncertainty in RDS. We prove the consistency of our method under some conditions and investigate its finite sample performance, through a simulation study, under realistic RDS sampling assumptions.
\end{abstract}

\noindent%
\vfill

\spacingset{1.45} 
\section{Introduction}
\label{sec:intro}

	Respondent-Driven Sampling (RDS) is a variant of link-tracing sampling that relies on social connections to reach members of `hard-to-reach' populations \citep{Hec97}. The RDS recruitment process starts off with the selection of initial \textit{seed} participants and runs through a number of recruitment \textit{waves} within which each selected individual is given a fixed number of \textit{coupons} and asked to recruit peers into the study.

	Inference for RDS has primarily focused on estimating population means and proportions. There is a rich literature on inference for population prevalence of viral diseases (\citealt{Gile10, Gile15b, Vol08}) and risk behaviors (\citealt{Mal08, Johnston2010TheAO, Joh08, Hec02}) among stigmatized populations. The vast majority of these studies use two classical and widely used approaches to inference for RDS, which rely on approximations to the true (and unknown) RDS sampling process. \cite{Vol08} approximated the RDS sampling process as a random walk on the nodes of an undirected graph and treated RDS samples as independent draws from its stationary distribution. The resulting inclusion probabilities are used to compute design weights known as RDS-II weights, applied to yield Horvitz-Thompson type estimators along with model-based estimates of uncertainty. This approach is  implemented in the \textsf{R} package \textsf{RDS}. Alternative estimators include the RDS-I and successive sampling estimators (\citealt{Gile11, Hec97}), however these are less commonly used and so we focus only on RDS-II.
	
	There is a growing interest in the estimation of uncertainty for RDS \citep{spiller2018evaluating} as recent findings show that most variance estimators (greatly) underestimate variability \citep{goel2010assessing}, thus yielding confidence intervals with coverage rates that are below expected nominal values. In fact, the widely used bootstrap method for RDS \citep{Salganik2006VarianceED}, which is implemented in \textsf{R} through the package \textsf{RDS}, assumes a first-order Markov dependence on categories defined by a variable of interest. The method vastly ignores the branching structure of RDS, which leads to an underestimation of uncertainty \citep{Gile18}. Further, the method is infeasible in regression settings because it does not deal with multiple variables at the same time. As an alternative variance estimation strategy, \cite{baraff2016estimating} proposed the \textit{tree bootstrap} method based on resampling the RDS recruitment tree. The method generates bootstrap trees from the observed recruitment tree in a hierarchical fashion. First, they sample with (or without) replacement from the initial seed participants. Then, they resample from each of the selected seeds' recruits. In the third level of the procedure, they resample from the second-level recruits' recruits. The sampling process continues until there are no recruits remaining. By mimicking the branching structure of the recruitment tree, this method aims to take into account the underlying network structure of the RDS sample, which other existing methods generally fail to achieve. Through simulation studies, \cite{baraff2016estimating} empirically demonstrated that the tree bootstrap method outperforms existing bootstrap methods, and yield confidence intervals with coverage rates at or above expected nominal values. Moreover, being the only known RDS bootstrap method for which resampling relies on the structure of the observed network and not on respondents' attributes, the tree bootstrap can yield estimates of any number of attributes from a single bootstrap sample, making it more computationally efficient than existing methods. However, recent findings suggest that this method can severely overestimate uncertainty \citep{Gile18} so that the cost of covering at or above the nominal level comes at a significant cost in terms of power.
	
	In this paper, we propose a bootstrap method for quantifying uncertainty in RDS that relies on the structure of the partially observed RDS network. The method is based on resampling recruited individuals and their observed \textit{neighbours},  i.e.~individuals with whom they are directly connected within the RDS tree. The paper is organized as follows. In Section \ref{sec:methods}, we present RDS as a sampling design with an underlying network structure. We then present the neighbourhood bootstrap method and prove its consistency under some regularity conditions. In Section \ref{sec:simulations}, we compare our method to the tree bootstrap method via a simulation study. In Section \ref{sec: casestudy}, we apply our method to a dataset collected through a cross-sectional study of gay, bisexual, and other men who have sex with men conducted in Montreal, Canada via  RDS.

\section{Methods}\label{sec:methods}
\subsection{Sampling design and assumptions}\label{sec:assum}
Consider a finite population of $N$ individuals. We assume that the individuals in the population are connected with social ties, or through a \textit{network}, represented by a \textit{graph} $G=(V, E)$, where $V$ represents the set of vertices, nodes or individuals, and $E$ represents the set of edges or social ties. An edge will be denoted by the ordered pair $(i, j)$, $i, j \in V$. For a \textit{directed} graph, ties are directed from one node to another, so $(i, j) \in E$ does not imply $(j, i) \in E$; for an \textit{undirected} graph, $(i, j) \in E$ imply $(j, i) \in E$.

We assume that the population network is an undirected, finite and connected graph. This implies that $(i)$ social connections are reciprocal and $(ii)$ an individual within the network can reach another individual through a finite set of connections. Let $a_{ij}$ be an indicator of social relationships between nodes $i$ and $j$ such that $a_{ij}=1$ in the presence of an edge, $a_{ij}=0$ otherwise, with $a_{ii}=0$. Note that $a_{ij}=a_{ji}$ under the assumption of reciprocal ties. We define $d_i=\sum_{j=1}^N a_{ij}$ as the \textit{degree} of node $i$, or the number of edges connected to that node.

Consider an RDS process that takes place within the population network and progresses across individuals' social ties. Let $G_T=(V_T, E_T)$ be the recruitment graph, or recruitment \textit{tree}, where $V_T\subseteq V$ is the set of recruited individuals, with $|V_T|=n$, such that $(i, j) \in E_T$ if individual $i$ recruited individual $j$, and $E_T \subseteq E$ is the set of recruiter-recruit relationships. This is a directed graph since relationships are directed from recruiter to recruit. We assume that the RDS process is a Markov process $\{X_u: u\in V_T\}$ on $G$ indexed by $G_T$ such that:
$$
\mbox{P}(X_w=j|X_v=i, X_u=k)=\mbox{P}(X_w=j|X_v=i)=P_{ij},
$$
where $(v, w)\in E_T$, $(u, w) \notin E_T$, and $P_{ij}\propto d_i$ is the $(i,j)-th$ element of the transition matrix $P \in \left[0, 1\right]^N$. We assume that the process begins at the stationary distribution of $P$. This is typically a $(G_T, P)$-walk on the population graph $G$ \citep{Ben94}.

\subsection{The RDS-II estimator}
 Let $Z_i \in \{0,1\}$ be a two-valued variable of interest indicating the presence or absence of an attribute for the ith individual, $i=1, \dots, N$. The target parameter is the population average 
$$
\mu=\frac{1}{N}\sum_{i=1}^N Z_i.
$$
An unbiased estimator of $\mu$ is the inverse probability weighted (IPW) estimator
$$
\hat\mu_{IPW}=\frac{1}{n}\sum_{u \in V_T} \frac{Z_u}{\pi_u},
$$
 where
$$
\pi_{u}=\frac{d_u}{\sum_{i=1}^N d_i}.
$$
The total degree $\sum_{i=1}^N d_i$ or the average degree $\sum_{i=1}^N d_i/N$ are usually unknown. Estimating the latter quantity with the mean degree $(n^{-1}\sum_{u\in V_T}d^{-1}_u)^{-1}$ yields the Volz–Heckathorn (VH) estimator for the population average \citep{Vol08}
$$
\hat\mu_{VH}=\frac{ \sum_{u\in V_T} Z_u\hat\pi_u^{-1}}{\sum_{u\in V_T} \hat\pi_u^{-1}},
$$
where
$$
\hat\pi_{u}=\frac{d_u}{n}\sum_{u\in V_T}\frac{1}{d_u}.
$$
In the next section, we propose a novel bootstrap variance estimator for $\hat\mu_{VH}$ and prove its consistency under some regularity conditions.

\section{A novel bootstrap variance estimator}\label{sec:bootmethods}
\subsection{The neighbourhood bootstrap}
	\cite{baraff2016estimating} proposed a variance estimator for $\hat{\mu}_{VH}$ based on resampling the RDS seed-induced trees. The method is described as follows. First, we sample with (or without) replacement from the seeds. Then, we resample from each of the seeds' recruits. In the third level of sampling, we resample from the second-level recruits' recruits. The process continues until there are no more recruits from which to sample. This method aims to take into account the underlying network structure of the RDS sample by mimicking the branching structure of the observed tree.

\begin{figure}[H]
\begin{center}
\includegraphics[scale=1.0]{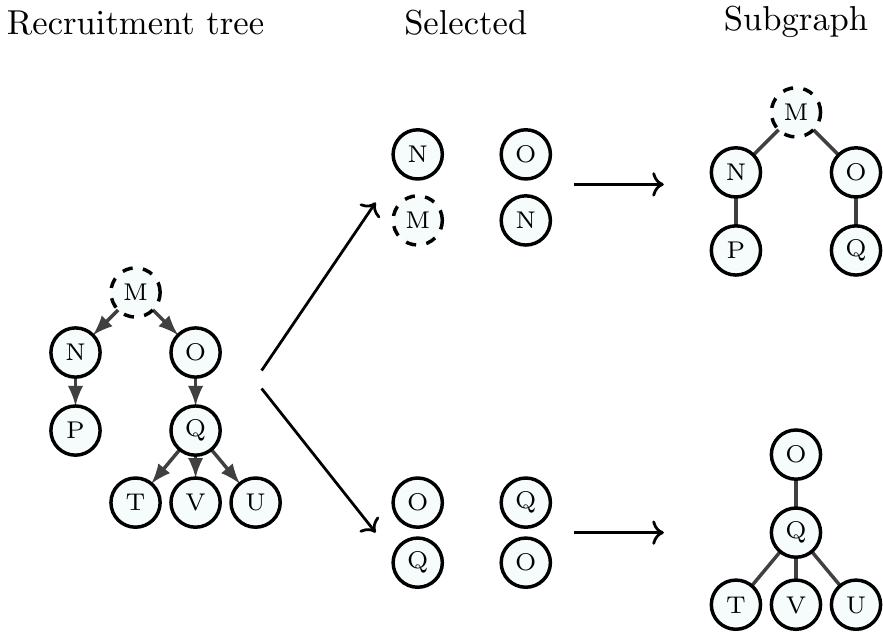}
\caption{Illustration of the neighbourhood bootstrap method with seed (dashed circle) labelled M, and subsequent participants N, O, ...V, U. The two arrows that point to the sampled nodes represent two different bootstrap resamples. We sampled $n_r=4$ participants in each bootstrap resample, where $n_r$ is the number of recruiters within the recruitment tree.}
\label{fig:dispopnetRDS}
\end{center}
\end{figure}

	We propose a bootstrap method based on sequentially resampling recruited individuals and their observed neighbours. First, we randomly select, with replacement, $n_r$ individuals, where $n_r$ is the number of recruiters within the resampled RDS tree. We then include the recruits of all selected individuals in the bootstrap sample. The network component of the resampled dataset is the subgraph induced by the recruits of the selected individuals. This is illustrated in Figure \ref{fig:dispopnetRDS}.

 The neighbourhood bootstrap method is designed with the hope that the sampling distribution of the RDS-II estimator captures the `local' neighbourhood structure of the network without mimicking the full branching structure of the RDS tree. This is motivated by the fact that the RDS tree is a partially observed network of unknown underlying dependence structure. Moreover, network clustering parameters such as \textit{homophily}, or the tendency for individuals with similar traits to share social connections, cannot be consistently estimated given the RDS tree alone (\citealt{Crawford17,shalizi2013consistency}).

\subsection{Consistency of the neighbourhood bootstrap}
In this section, we establish the consistency of the neighbourhood bootstrap, and that of corresponding confidence intervals. Consider an RDS design in which $s$ seeds are selected and given $c$ coupons each to distribute among their neighbours. Suppose each seed, and their selected neighbours, successfully recruit exactly $c$ other neighbours into the study through $h$ recruitment waves. In this setting, each seed gives rise to a seed-rooted tree of $r=(c^{h+1}-1)/(c-1)$ recruits for an RDS sample size of $n=sr$. 

Let $\bar{Z}_n=n^{-1}\sum_{u\in V_T}Z_u$ be the RDS sample average. This is an unbiased estimator of $\mu_0=\sum_{i=1}^N Z_i\pi_i$ since the walk is assumed to begin at the stationary distribution. For any RDS dataset $Z_1, \dots, Z_n$, let $\bar{Z}_n^{*}=n^{-1}\sum_{u\in V_T} Z^{*}_u$ be the bootstrap average, where $Z^{*}_1, \dots, Z^{*}_n$ is a bootstrap sample. Let $\mbox{E}_{RDS}[\cdot]$ ($\mbox{Var}_{RDS}[\cdot]$) and $\mbox{E}_{*}[\cdot]$ ($\mbox{Var}_{*}[\cdot]$) denote the expectation (variance) with respect to the RDS process and the neighbourhood bootstrap method, respectively. First, we show the uniform convergence of the distribution of the bootstrap approximation $\bar{Z}_n^{*}-\bar{Z}_n$ to that of $\bar{Z}_n-\mu$, under the condition assumed in the next theorem. This suffices to establish the consistency of the corresponding bootstrap confidence intervals. We then use this result to further show the uniform convergence of the distribution of $\hat\mu_{VH}^{*}-\hat\mu_{VH}$ to that of $\hat\mu_{VH}-\mu)$, where $\hat\mu_{VH}^{*}$ is the VH estimator of the bootstrap sample $Z^{*}_1, \dots, Z^{*}_n$.

We consider the asymptotic distribution of $\bar{Z}_n^{*}-\bar{Z}_n$ as $n\rightarrow \infty$. Since we are operating in a finite population setting, with $n\leq N$, we also require $N\rightarrow \infty$. This can be achieved by embedding our population in a sequence of increasing finite populations. We let $n\rightarrow \infty$, $N-n \rightarrow \infty$, $\lim_{N \to \infty} \sum_{i=1}^N Z_i/N=\mu$, and $\lim_{N \to \infty} \mbox{Var}(\sqrt{N}(\bar{Z}_N-\mu))=\sigma^2<\infty$, where $\mbox{Var}[\cdot]$ denotes the variance with respect to the infinite population model. We also assume that $d_1, \dots, d_N$ are i.i.d random variables taking values in $\mathbb{N}$. Finally, we assume the following rate requirement: $r=o(\sqrt{n})$. This implies that $s\rightarrow \infty$ as $n\rightarrow \infty$. 

\begin{theo}{\textbf{(Consistency).}}\label{theo:consis}
Consider the following condition:
\begin{enumerate}
\item $\sqrt{n}(\bar{Z}_n-\mu) \xrightarrow{d} N(0, \sigma^2)$. This central limit theorem holds under the assumption that the design effect of the estimator is finite, i.e. when $c<1/\lambda^2$, where $\lambda$ is the second eigenvalue of the transition matrix $P$ \citep{Rohe2019ACT}.
\end{enumerate}
If the above condition holds, then
\begin{equation}
\adjustlimits\sup_{x \in R} |\mbox{F}_{\bar{Z}_n^{*}-\bar{Z}_n}(x)-\mbox{F}_{\bar{Z}_n-\mu}(x)| \xrightarrow{a.s.} 0,
\end{equation}
where $\mbox{F}_{\bar{Z}_n^{*}-\bar{Z}_n}$ represents the c.d.f. of $\bar{Z}_n^{*}-\bar{Z}_n$ given $Z_1, \dots, Z_n$, and $\mbox{F}_{\bar{Z}_n-\mu}$ represents the c.d.f. of $\bar{Z}_n-\mu$.
\end{theo}
The proof is given in the Appendix. 
\begin{coro}\label{coro:IPW}
One has:
\begin{equation*}
\adjustlimits\sup_{x \in R} |\mbox{F}_{\hat\mu_{IPW}^{*}-\hat\mu_{IPW}}(x)-\mbox{F}_{\hat\mu_{IPW}-\mu}(x)| \xrightarrow{a.s.} 0,
\end{equation*}
where $\hat\mu_{IPW}^{*}$ is the IPW estimator of the bootstrap sample $Z^{*}_1, \dots, Z^{*}_n$.
\end{coro}
The proof follows by substituting $Z_u^{\pi}=N^{-1}\pi^{-1}_uZ_u$ for $Z_u$ in Theorem \ref{theo:consis}. The next Corollary establishes the consistency of bootstrap confidence intervals for the VH estimator.

\begin{coro}\label{lemm:VH}
One has:
\begin{equation*}
\adjustlimits\sup_{x \in R} |\mbox{F}_{\hat\mu_{VH}^{*}-\hat\mu_{VH}}(x)-\mbox{F}_{\hat\mu_{VH}-\mu}(x)| \xrightarrow{a.s.} 0.
\end{equation*}
\end{coro}
The proof is given in the Appendix. The consistency of the neighbourhood bootstrap method has been established under the assumptions that each individual successfully distributes all coupons. Further, the approximation of the RDS process as a $(G_T, P)$-walk on the population graph implies that sampling is done with replacement. These assumptions are not realistic in most real-world RDS studies. In the next section, we investigate the finite sample performance of our method (the bias of the bootstrap variance estimator and the coverage of the bootstrap confidence interval) when these assumptions are relaxed.

\section{Monte Carlo simulations}\label{sec:simulations}
We conducted a simulation study using a real-world dataset of networked individuals from the Colorado Springs Project 90 study \citep{klovdahl1994social}. From 1988 through 1992, data on 13 demographic characteristics and risk behaviors were collected on people who inject drugs and their associates (sexual and needle-sharing partners) and sex workers and their (paying and non paying) partners. Respondents listed their personal network of contacts within the community, resulting in the construction of the full network of social relationships. Around 600 respondents contributed in building a social network of 5493 individuals and 21644 connections, distributed among 125 connected clusters. The single largest cluster connected 4430 individuals through 18407 ties. The goal of the study was to investigate the effects of the network structure on the dynamics of HIV transmission in a community of high-risk heterosexuals. This dataset was also used by \cite{baraff2016estimating} while comparing various bootstrap methods for RDS.

We considered the largest connected cluster of the network as the population network and simulated RDS samples using the same setting as \cite{baraff2016estimating}. First, we randomly sampled, without replacement, $s=10$ seeds with probability proportional to degree. Each sampled individual recruited uniformly, without replacement, up to $c=3$ individuals, with probabilities of recruiting $0$, $1$, $2$ and $3$ individuals set at $1/3$, $1/6$, $1/6$ and $1/3$, respectively. The process was stopped once the sample size reached the desired levels; the sample size was set to $n=500$, $n=800$ and $n=1000$. We simulated 1000 RDS samples from the Project 90 data. For each simulated sample, we used the tree bootstrap and the neighbourhood bootstrap methods to compute $95\%$ percentile confidence intervals for the population proportions of the 12 demographic characteristics shown in Table \ref{table:simulation_CVGE}. The results are presented in the next section.

\subsection{Coverage and mean interval width}
\begin{figure}[H]
\begin{center}
\includegraphics[scale=.65]{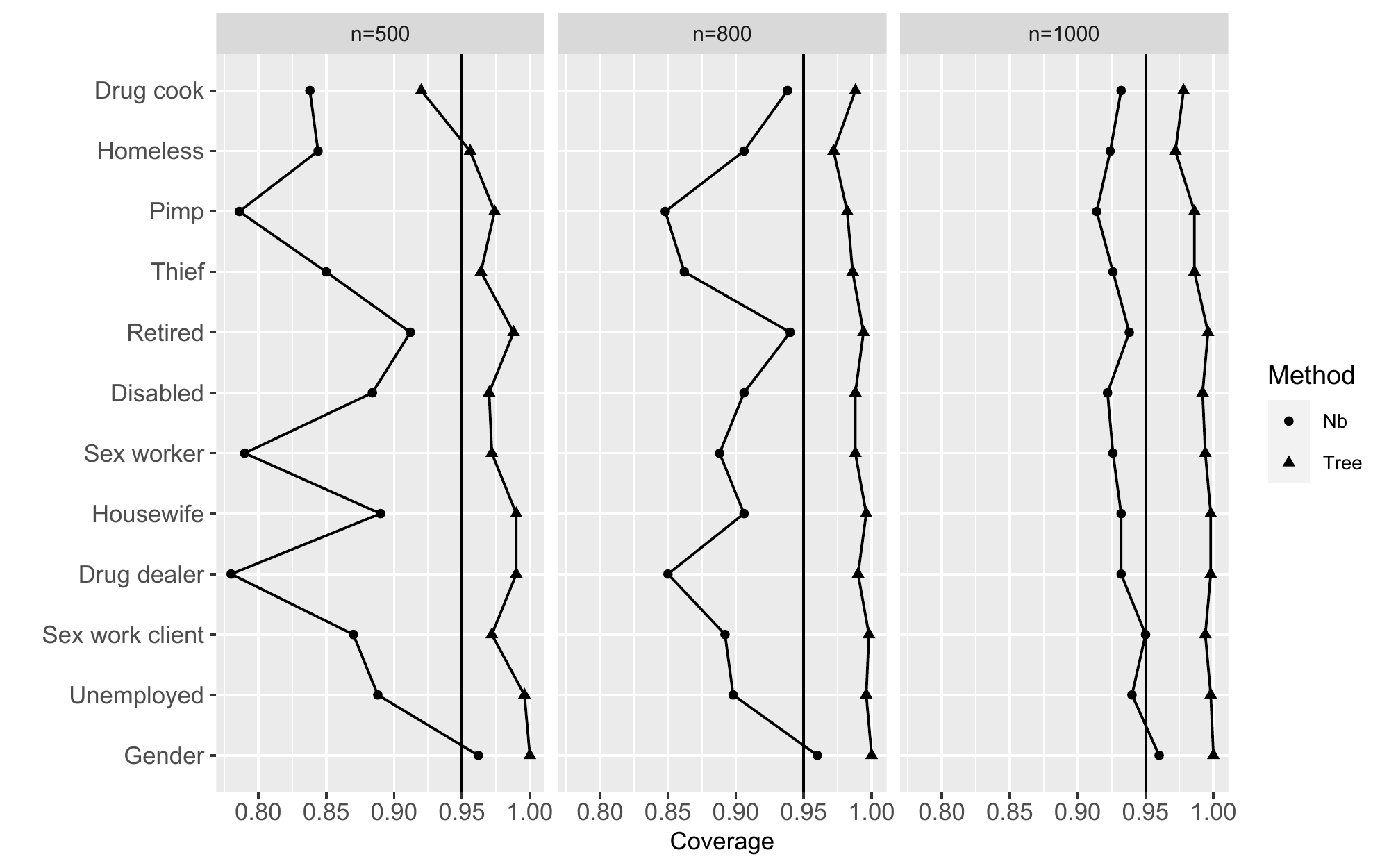}
\end{center}
\caption{Project 90 - Coverage probabilities of the $95\%$ confidence intervals obtained through the neighbourhood bootstrap (Nb) and the tree bootstrap (Tree) methods when sampling is done without replacement.}
\label{fig:covg}
\end{figure}

Figure \ref{fig:covg} displays the coverages of the $95\%$ (percentile) confidence intervals obtained through the tree bootstrap and the neighbourhood bootstrap methods (coverages of the $80\%$ confidence intervals are displayed in the Appendix). The results show that the tree bootstrap method yields coverages that are above and closer, in some cases, to the nominal values than the neighbourhood bootstrap method when $n=500$. However, as the sample size increases, the neighbourhood method yields coverages that are closer to the nominal values than the tree method across all attributes.

\begin{figure}[H]
\begin{center}
\includegraphics[scale=.65]{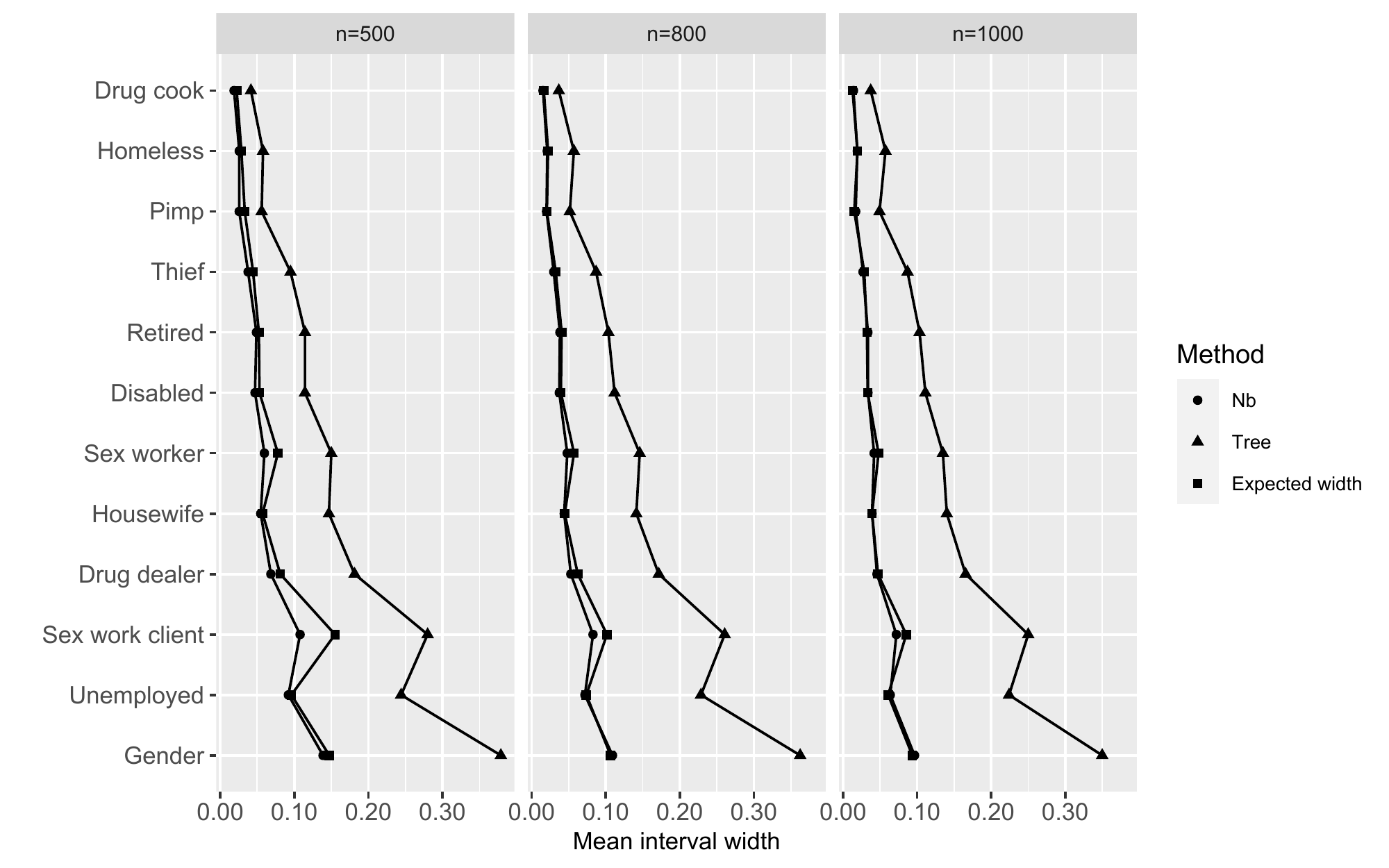}
\end{center}
\caption{Project 90 - Mean interval width of the $95\%$ confidence intervals obtained through the neighbourhood bootstrap (Nb) and the tree bootstrap (Tree) methods when sampling is done without replacement.}
\label{fig:Width95}
\end{figure}

We further compare the mean widths of the 95\% and 80\% (see Appendix) confidence intervals obtained through both methods by generating 5000 RDS samples and computing the expected widths of the intervals using the sampling distribution of the RDS-II estimators for all attributes. The results, displayed in Figure \ref{fig:Width95}, show that the average widths of our method were far closer to the expected widths than those from the tree bootstrap method across all attributes.

\subsection{Biases of the variance estimators}
We also investigated the accuracy of the competing bootstrap variance estimators. For each replication of the simulation, we ran 1000 replications of the tree bootstrap and the neighbourhood bootstrap procedures to compute the bootstrap variance for $\hat \mu$, $v(\hat \mu)$. For each attribute, the true variance is the mean squared error (MSE) of the corresponding RDS-II estimator. The relative bias for each variance estimator is computed as follows:
$$
\mbox{RB}\left[\mbox{E}\{ v(\hat \mu)\}\right]=\left[\mbox{E}\{ v(\hat \mu)\}-\mbox{MSE}(\hat \mu)\right]/\mbox{MSE}(\hat \mu),
$$
where $\mbox{MSE}(\hat \mu)=\{\sum_i (\hat{\mu}_{i}-\mu)^{2}/1000\}$, where $i$ indexes the bootstrap replication. Table \ref{table:simulation_CVGE} reports the relatives biases for the tree bootstrap and the neighbourhood bootstrap variance estimators across 12 attributes.

 The results show that the tree bootstrap variance estimator overestimates variability (relative bias greater than $100\%$) across all attributes and sample sizes. The neighbourhood bootstrap yields a small to negligible bias for most attributes across all sample sizes; the corresponding bias for the variance estimator is negative for $n=500$ and $n=800$, explaining coverage probabilities that are below expected nominal values in most cases.

\begin{table}[H]
\caption{Project 90 - Relative biases (in percentages) for the tree bootstrap (Tree) and the neighbourhood bootstrap (Nb) variance estimators for the RDS-II estimates of proportions for 12 variables.}
\begin{center}
\setlength\extrarowheight{-2pt}
\small
\begin{tabular}{ lcc cccccc} \hline
\multicolumn{3}{r}{$n=500$}&&&{$n=800$}&&&{$n=1000$}\\
    \cline{2-3}     \cline{5-6} \cline{8-9}
 {Variables}& {Tree}&{Nb}& & {Tree}&{Nb}&&{Tree}&{Nb}\\ \hline
 Gender  & 5.95 & 0.08  & &9.76&0.04 & &11.58&0.02  \\
 Sex worker  &  3.57 &  -0.37  & &7.56&-0.24 &&11.29&-0.15  \\
 Pimp &  3.57 &  -0.48   & &9.71& 0.01&&16.10&0.14 \\
 Sex work client & 3.00  & -0.26  & &6.65&-0.32&&10.72&0.02\\
 Drug dealer &  4.78 &-0.26    & &8.80& -0.13& &13.76&-0.06\\
 Drug cook &  7.02 &  0.14  & &10.54& 0.19& &12.35&0.28\\
 Thief  & 6.30  &  -0.09     & &11.30& -0.08&  &14.74&0.00 \\
 Retired  &  5.89&  -0.08 & &9.59&0.07 &  &12.80&0.09\\
 Housewife & 6.46  & -0.13 & &11.03&-0.04 & &14.75&0.28 \\
 Disabled &  6.55 &  -0.14 & &10.25& -0.03&  &11.85&0.08\\
 Unemployed & 6.19 &  0.12    & &10.55&0.04 & &14.06&0.36 \\
 Homeless & 6.40  &  -0.16 & &10.90& -0.07&  &16.29&0.16 \\
\hline
\end{tabular}

\end{center}
\label{table:simulation_CVGE}
\end{table}

	These results show that while the tree bootstrap method consistently provides coverage at or above the nominal level, it also provides variance estimators that greatly overestimate variability across all sample sizes, implying a significant cost in terms of power. Our method provides coverage at or slightly below the nominal level with increasing sample size, while yielding far less biased variance estimators than those obtained through the tree bootstrap method.

\section{Case Study}\label{sec: casestudy}
	We applied our method to an RDS dataset collected through the Engage study, a national cross-sectional study conducted in Montreal, Toronto and Vancouver, with the goal of determining the individual, social and community-level factors that impact transmission of HIV and sexually transmitted infections among gay, bisexual, and other men who have sex with men (GBM) (\citealt{DRSP19,Doyle20}). In Montreal, the recruitment started with the selection of $27$ participants, chosen to be as heterogeneous as possible with respect to the diversity (e.g., HIV status, ethnicity) of the GBM community. Each selected participant received $6$ uniquely identified coupons to recruit their peers into the study. The Montreal arm of the study was conducted from February 2017 through June 2018, and recruited a total of 1168 men.

About $33\%$ of participants were aged less than 30, around $70\%$ were born in Canada, around $30\%$ had a diploma that is lower than college's, and about 3 out of five participants earned less than \$30,000 CAD annually. When it comes to risk behaviors, around $13\%$ of participants declared using crack cocaine, about $6\%$ declared using a syringe used by someone else, and $17\%$ of participants were living with HIV.

	For each attribute, we computed estimates for the population prevalence using the self-reported network degrees. Further, we used the neighbourhood method and the tree bootstrap method to compute $95\%$ (percentile) confidence intervals for the estimates. The tree bootstrap yields larger standard errors and wider confidence intervals than the neighbourhood method. As highlighted in the simulation study, we expect that the neighbourhood methods yield narrower confidence intervals with coverage probabilities that are closer to expected nominal values than those of the tree bootstrap.

\begin{table}[H]
\caption{Estimates (Est.) of proportions of various socio-demographic, behavioural and clinical factors among gays, bisexuals and men who have sex with men in Montreal, Canada. Standard errors (SE) and 95\% confidence intervals (CI) were computed using the tree bootstrap method (Tree) and the neighbourhood bootstrap methods (Nb).}
\begin{center}
\small
\begin{tabular}{ ccc ccccc} \hline
\multicolumn{3}{r}{Tree}&&&&{Nb}\\
    \cline{2-4}     \cline{6-8}
 {Variables}& {Est.}&{SE}&{CI}& & {Est.}&{SE}&{CI}\\ \hline
 Age$<$30  & 0.36 & 0.12 &  [0.16, 0.58]  & &0.36&0.02 &[0.32, 0.40] \\
 Born in Canada &  0.65 &  0.09 &  [0.49, 0.82]  & &0.65&0.02 &[0.61, 0.69] \\
Diploma$<$college &  0.35 &  0.10 & [0.18, 0.54]   & &0.35& 0.02&[0.31, 0.39]\\
Crack use &0.11 & 0.06 & [0.04, 0.25]   & &0.11&0.02 &[0.08, 0.14]\\
Syringe use &  0.05 &0.03  & [0, 0.12]   & &0.05& 0.01&[0.04, 0.07]\\
HIV-positive &  0.14&  0.06 &  [0.05, 0.28]  & &0.14& 0.01&[0.12, 0.17]\\
\hline
\end{tabular}
\end{center}
\label{table:Engage_CVGE}
\end{table}

\section{Conclusion}
Recent findings suggest that the tree bootstrap method, although practical in capturing the variability in RDS estimates, overestimates uncertainty and yield confidence intervals that are too wide in some cases. We proposed a bootstrap method, based on resampling recruited individuals and their neighbours within the RDS tree, that can reasonably capture the variability in the estimates. Our method produces confidence intervals that are narrower than those of the tree bootstrap, with coverage probabilities that are closer to the nominal values as the sample size increases in all simulations that we considered. Further, we empirically showed that the tree bootstrap greatly overestimates the variance associated with RDS estimates (relative biases greater than 100\%) while our neighbourhood bootstrap method yields less biased variance estimators across all sample sizes. 

We have implemented the neighbourhood bootstrap in an R package, \textsf{Neighboot}, available on CRAN \citep{Yau20neighboot}.

\appendix

\section*{Appendix}
\subsection*{Proof of Theorem \ref{theo:consis}}
Following Polya's theorem \citep{Kiefer1960ModernPT}, we only need to show the following:
\begin{equation*}
\sqrt{n}(\bar{Z}_n^{*}-\bar{Z}_n)\rightarrow N(0, \sigma^2).
\end{equation*}
We will establish this result by $(i)$ proving Proposition \ref{prop:unbias}, stated below, and by $(ii)$ using the Linderberg-Feller central limit theorem.
First, let $G_{T,j}$ be the $j$th seed-rooted tree, $j=1, \dots, s$. A neighbourhood bootstrap sample is obtained by randomly selecting with replacement $\ell=(c^h-1)/(c-1)$ recruiters from each seed-rooted tree and by including their recruits. Let $G_{T,jb}$ be the $b$th subgraph induced by the $b$th bootstrap sample within the $j$th seed-rooted tree, $b=1, \dots, S$, where $S=\ell^{\ell}$.

\begin{prop}\label{prop:unbias}
$n\mbox{Var}_{*}(\bar{Z}_n^{*}) \xrightarrow{a.s.} \sigma^2$  as $n \rightarrow \infty$.
\end{prop}
We first give the proof of Proposition \ref{prop:unbias}. Let $R^b_{u(j)}=1$ if $u \in V_{T,jb}$, and $R^b_{u(j)}=0$ otherwise. Since any individual will appear in the $b$th bootstrap sample if his/her recruiter is selected, it is clear that 
$$
\mbox{P}\left(R^b_{u(j)}=1 \right)=\frac{1}{r}.
$$
Further, 
$$\mbox{P}\left(R^b_{u(j)}=1, R^b_{v(j)}=1 \right)=\frac{1}{r}$$ if $u$ and $v$ share the same recruiter, and $$\mbox{P}\left(R^b_{u(j)}=1, R^b_{v(j)}=1 \right)=\frac{1}{r^2}$$ otherwise. For the $b$th bootstrap sample, the bootstrap average is given by 
\begin{eqnarray*}
\bar{Z}^{*}_n&=&\frac{1}{s}\sum_{j=1}^s \frac{1}{r}\sum_{u \in V_{T,jb}} Z_u.
\end{eqnarray*}
Thus,
\begin{eqnarray*}
\mbox{E}_{*}(\bar{Z}^{*}_n)&=&\frac{1}{s}\sum_{j=1}^s \frac{1}{r} \sum_{u \in V_{T,jb}} \mbox{E}_{*}(Z_u)\\
&=&\frac{1}{s}\sum_{j=1}^s \frac{1}{r} \sum_{u \in V_{T,jb}} \sum_{v \in V_{T,j}}Z_u\frac{1}{r} \\
&=& \frac{1}{s}\sum_{j=1}^s \frac{1}{r}\sum_{v \in V_{T,j}} Z_v \\
&=&\bar{Z}_n.
\end{eqnarray*}
Now, let $V_{T,b}=\cup_{j=1}^s V_{T,jb}$ for $b=1,\dots, S$. One has:
\begin{eqnarray}\label{eq:varZ1}
\nonumber \mbox{Var}_{*}(\bar{Z}^{*}_n)&=&\frac{1}{n^2}\mbox{Var}_{*}(\sum_{u \in V_{T,b}}Z_u)\\
&=&\frac{1}{n^2}\left\lbrace  \sum_{u \in V_{T,b}} \mbox{Var}_{*}(Z_u) + \sum_{u\neq v} \mbox{cov}_{*}(Z_u, Z_v) \right\rbrace.
\end{eqnarray}
One has
\begin{eqnarray*}
\sum_{u \in V_{T,b}} \mbox{Var}_{*}(Z_u)&=&n\sum_{v \in V_T} \{Z_v-\mbox{E}_{*}(Z_v)\}^2\frac{1}{n}\\
&=&\sum_{v \in V_T} (Z_v-\bar{Z}_n)^2.
\end{eqnarray*}
Further, straightforward developments lead to
\begin{eqnarray*}
\sum_{u\neq v} \mbox{cov}_{*}(Z_u, Z_v)=\frac{n-1}{n^2}\sum_{j=1}^s \sum_{u \in V_{T,j}}  Z^2_u+ \frac{n-1}{n^2}\sum_{j=1}^s \sum_{u \in \mathcal{N}^j(v)}Z_uZ_v, 
\end{eqnarray*}
where $\mathcal{N}^j(v)$ represent individuals who share the same recruiter with individual $v$ within the $j$th tree. It follows that
\begin{eqnarray*}
n\mbox{Var}_{*}(\bar{Z}^{*}_n)&=&\frac{1}{n}\sum_{u \in V_T} (Z_u-\bar{Z}_n)^2+\frac{n-1}{n^3}\sum_{u \in V_{T}}  Z^2_u+\frac{n-1}{n^3}\sum_{u \in \mathcal{N}(v)}Z_uZ_v.
\end{eqnarray*}
It is clear that $n^{-1}\sum_{u \in V_{T}}  Z^2_u=\mathcal{O}_P(1)$ and $n^{-1}\sum_{u \neq v}  Z_u Z_v=\mathcal{O}_P(1)$. It follows that
$$
\lim_{n \to \infty} n\mbox{Var}_{*}(\bar{Z}^{*}_n)=\lim_{n \to \infty} \frac{1}{n}\sum_{u \in V_T} (Z_u-\bar{Z}_n)^2=\sigma^2.
$$

We now show that 
\begin{eqnarray}\label{eq:CLT1}
\frac{\bar{Z}_n^{*}-\bar{Z}_n}{\sqrt{\mbox{Var}(\bar{Z}_n^{*})}} \xrightarrow{d} N(0, 1).
\end{eqnarray}
Let
\begin{eqnarray*}
Y_{nj}&=&\frac{1}{n}\left(\sum_{u \in V_{T,jb}}  Z_u -\sum_{u \in V_{T,j}}  Z_u \right),\\
T_{n}&=&\sum_{j=1}^s Y_{nj},\\
s^2_n&=&\mbox{Var}_{*}(T_n)=\mbox{Var}_{*}(\bar{Z}^{*}_n).
\end{eqnarray*}
One has $\mbox{E}_{*}(Y_{nj})=0$ and $\mbox{Var}_{*}(Y_{nj})<\infty$.
We prove (\ref{eq:CLT1}) by verifying the following Linderberg condition:
\begin{eqnarray}\label{eq:CLT2}
\frac{1}{s^2_n}\sum_{j=1}^s \mbox{E}_{*}\left\lbrace Y^2_{nj} I\left(|Y_{nj}|\geq \epsilon s_n \right)  \right\rbrace  \rightarrow 0 \,\, \text{as}\,\, n \rightarrow \infty,
\end{eqnarray}
for every $\epsilon>0$. Since $|Y_{nj}|\leq r/n$, (\ref{eq:CLT2}) is zero whenever 
$$
\frac{n}{r}\epsilon s_n>1.
$$
This is achieved when $n^2r^{-2} s^2_n \rightarrow \infty$ as $n\rightarrow \infty$. One has:
\begin{eqnarray*}
\frac{n^2}{r^2} s^2_n&=&\frac{n}{r^2}\times n\mbox{Var}_{*}(\bar{Z}^{*}_n).
\end{eqnarray*}
Since $n\mbox{Var}_{*}(\bar{Z}^{*}_n)=\mathcal{O}(1)$ and $r^2=o(n)$, 
$$
\frac{n}{r^2}\times n\mbox{Var}_{*}(\bar{Z}^{*}_n) \rightarrow \infty\,\, \text{as}\,\, n \rightarrow \infty,
$$
which completes the proof.

\subsection*{Proof of Corollary \ref{lemm:VH}}
We will show that the asymptotic distribution of $\hat\mu_{VH}-\mu$ is identical to that of $\hat\mu_{IPW}-\mu$. The same approach can be used to show that the asymptotic distribution of $\hat\mu_{VH}^{*}-\hat\mu_{VH}$ is identical to that of $\hat\mu_{IPW}^{*}-\hat\mu_{IPW}$. These results will complete the proof given Corollary \ref{coro:IPW}.

The quantity $\hat\mu_{VH}-\mu$ can be expressed as:
\begin{eqnarray}\label{eq:lemm1}
\hat\mu_{VH}-\mu&=&C(N, n, \{d_i\})\left(\hat\mu_{IPW}-\mu\right)+\left\lbrace C(N, n, \{d_i\})-1  \right\rbrace \mu,
\end{eqnarray}
where 
$$
C(N, n, \{d_i\})=\frac{n}{\sum_{u\in V_T}d_u^{-1}}\frac{N}{\sum_{i=1}^n d_i}.
$$
Let $d=\lim_{N \to \infty} \sum_{i=1}^N d_i/N $. This limit exists since $d_1, \dots, d_N$ are i.i.d random variables taking values in the set $\mathbb{N}$. Now, the limit of the harmonic mean of the observed degrees is
$$
\lim_{n \to \infty} \frac{1}{n}\sum_{u\in V_T}\frac{1}{d_u}=\frac{1}{d},
$$
using the SLLN. It follows that
\begin{equation*}
\lim_{n \to \infty} C(N, n, \{d_i\})=d\times \frac{1}{d}=1.
\end{equation*}
Combining this result with Slutsky's theorem in Equation (\ref{eq:lemm1}) shows that the limit distribution of $\hat\mu_{VH}-\mu$ is identical to that of $\hat\mu_{IPW}-\mu$. This completes the proof.

\subsection*{Coverage and mean interval width of the $80\%$ confidence intervals }
\begin{figure}[H]
\begin{center}
\includegraphics[scale=.65]{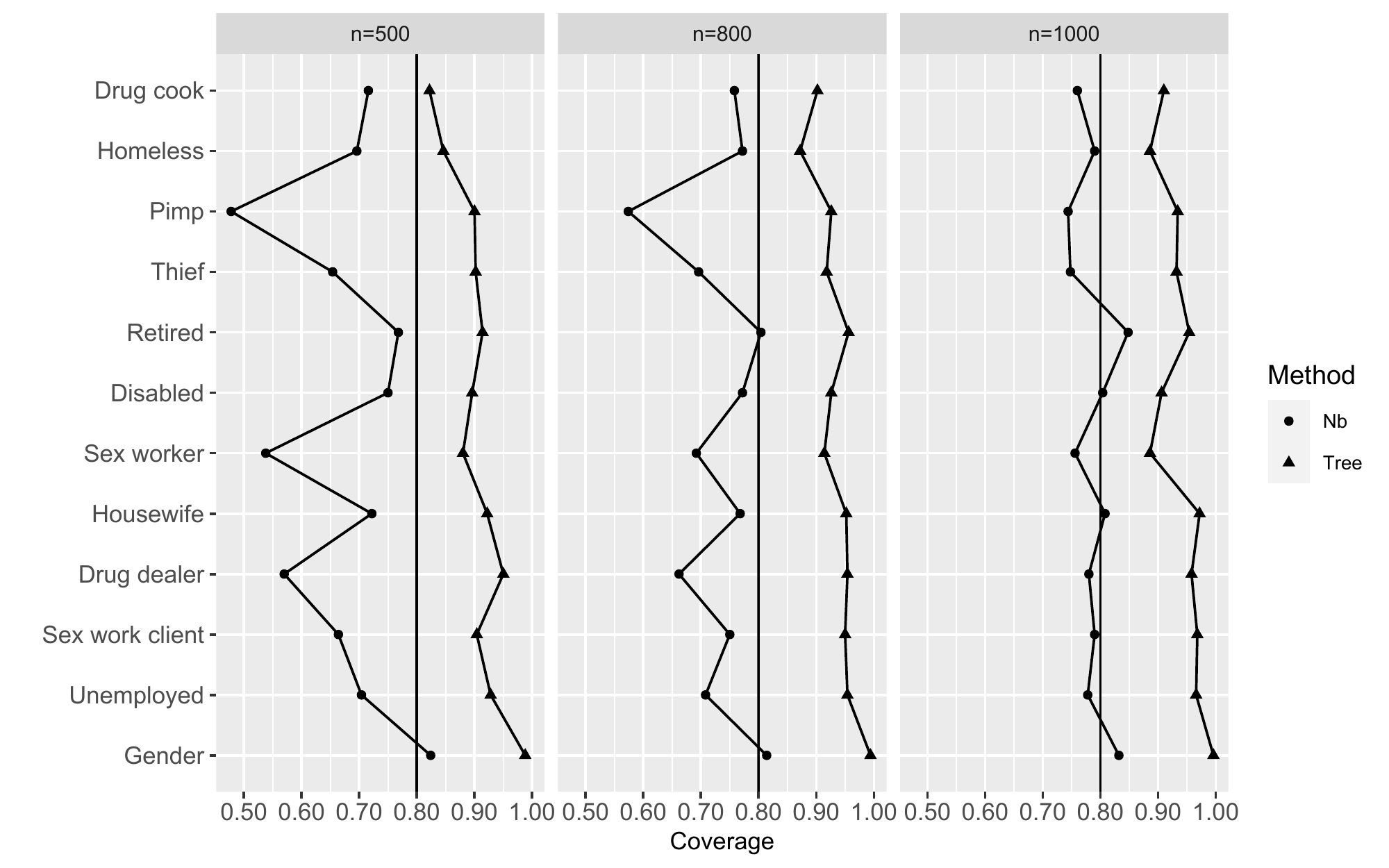}
\caption{Project 90 - Coverage probabilities of the $80\%$ confidence intervals obtained through the neighbourhood bootstrap (Nb) and the tree bootstrap (Tree) methods when sampling is done without replacement. RDS samples were drawn from the Project 90 network.}
\end{center}
\label{fig:covg80}
\end{figure}

\begin{figure}[H]
\begin{center}
\includegraphics[scale=.65]{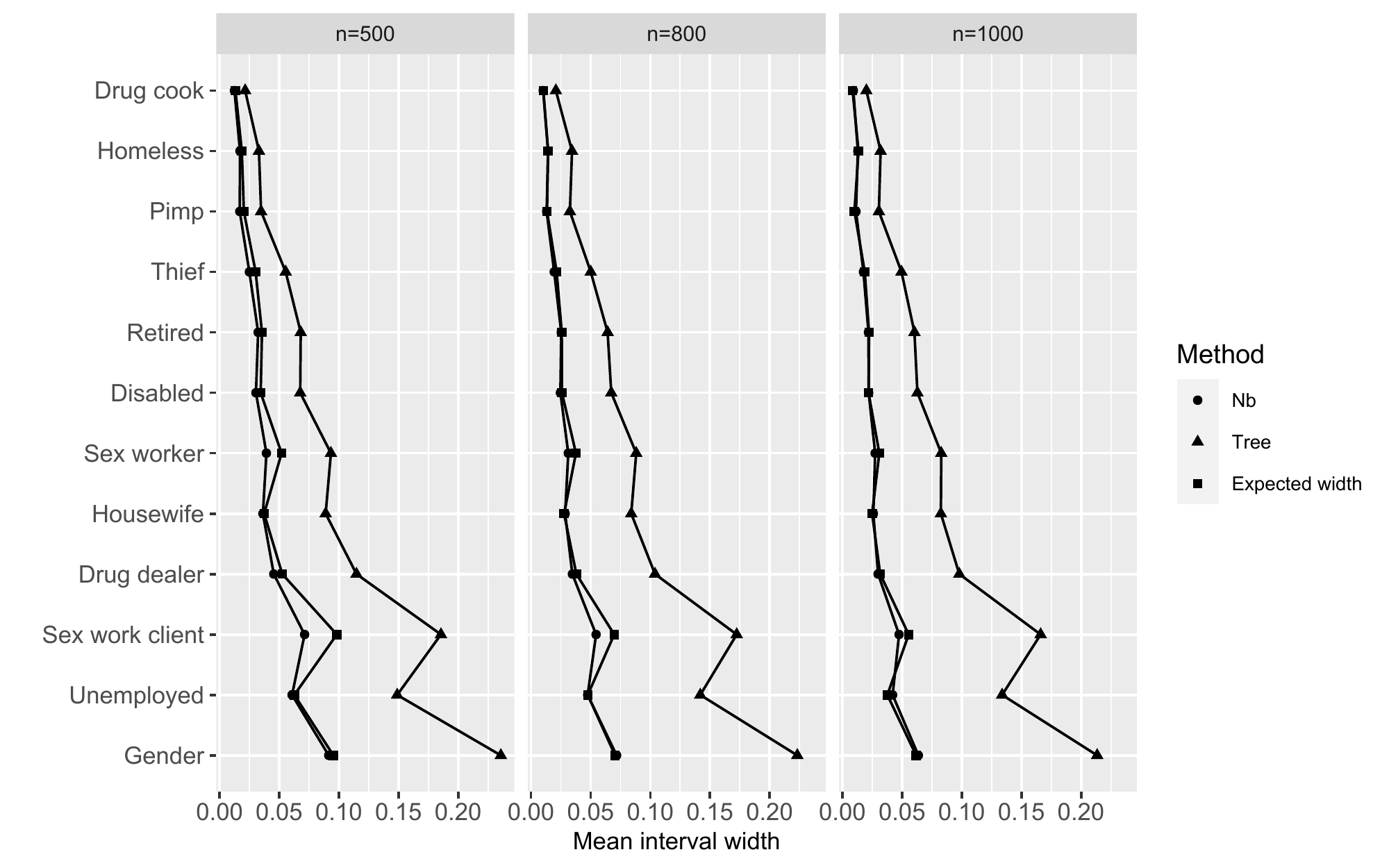}
\end{center}
\caption{Project 90 - Mean interval width of the $80\%$ confidence intervals obtained through the neighbourhood bootstrap (Nb) and the tree bootstrap (Tree) methods when sampling is done without replacement.}
\label{fig:Width80}
\end{figure}

\end{document}